\documentclass[sigconf]{acmart}

\AtBeginDocument{%
  }

\copyrightyear{2024} 
\acmYear{2024} 
\setcopyright{rightsretained} 
\acmConference[SIGCSE Virtual 2024]{Proceedings of the 2024 ACM Virtual Global Computing Education Conference V. 2}{December 5--8, 2024}{Virtual Event, NC, USA}
\acmBooktitle{Proceedings of the 2024 ACM Virtual Global Computing Education Conference V. 2 (SIGCSE Virtual 2024), December 5--8, 2024, Virtual Event, NC, USA}
\acmDOI{10.1145/3649409.3691079}
\acmISBN{979-8-4007-0604-2/24/12}

\begin{document}

\title[FSMIPR]{Finite State Machine with Input and Process Render}





 \author{Sierra Zoe Bennett-Manke}
  \email{sierrazoe.bennettmanke@westpoint.edu}
 \authornote{The first two authors contributed equally to this research.}
 \orcid{0009-0002-6681-8318}
 \affiliation{%
   \institution{United States Military Academy}
   \city{West Point}
   \state{New York}
   \country{USA}
   \postcode{10996}}

 \author{Sebastian Neumann}
  \email{sebastian.neumann@westpoint.edu}
  \orcid{0009-0000-9857-5117}
 \affiliation{
   \institution{United States Military Academy}
   \city{West Point}
   \state{New York}
   \country{USA}}

\author{Ryan E. Dougherty}
\email{ryan.dougherty@westpoint.edu}
\orcid{0000-0003-1739-1127}
\affiliation{%
  \institution{United States Military Academy}
  \streetaddress{601 Thayer Road}
  \city{West Point}
  \state{New York}
  \country{USA}
  \postcode{10996}
}

\renewcommand{\shortauthors}{Bennett-Manke, Neumann, and Dougherty}

\begin{abstract}
  Finite State Machines are a concept widely taught in undergraduate theory of computing courses. 
  Educators typically use tools with static representations of FSMs to help students visualize these objects and processes; however, all existing tools require manual editing by the instructor. 
  In this poster, we created an automatic visualization tool for FSMs that generates videos of FSM simulation, named Finite State Machine with Input and Process Render (FSMIPR).
  Educators can input any formal definition of an FSM and an input string, and FSMIPR generates an accompanying video of its simulation.
  We believe that FSMIPR will be beneficial to students who learn difficult computer theory concepts. 
  We conclude with future work currently in-progress with FSMIPR.
\end{abstract}

\ccsdesc[500]{Social and professional topics~Computing education}
\ccsdesc[300]{Theory of computation}

\keywords{theory of computing, finite automaton, manim, visualization, cs education}

\received{20 February 2007}
\received[revised]{12 March 2009}
\received[accepted]{5 June 2009}

\maketitle

\section{Introduction}

Finite state machines are a mathematical model of computation used to represent abstract states and the transitions between those states which are in response to some input. 
There are many different types of state machines, including Deterministic Finite Automata (DFAs).
These machines are essential in undergraduate Theory of Computing (ToC) classes as they can model different types of modern computers. 
Since these state machines have formal and technical definitions, students often have issues understanding their behavior.
Instructors often solve this problem with board visualizations of these machines as state diagrams with transitions.
This method is laborious and sometimes unhelpful as the instructor needs to layout the states and transitions in a logical way, which may be different depending on the machine's behavior.
Further, although using a static representation can be useful in the classroom, rendering a simulation of the machine by-hand takes valuable time, and updating the simulation's multiple data can be initially confusing for students.
There are existing FSM visualization tools but none solve all of the above challenges. 

In this poster we present Finite State Machine with Input and Process Render (FSMIPR) that automatically creates visualizations of state machine simulation on a given input string.
Our tool can be applied to any DFA with any input string.
The benefit of such a tool is that it is automatic, and both shows the DFA state diagram and its formal transition table.
This both saves instructor workload in making such a visualization and reduces potential instructor error.
Most importantly, it provides a bridge between the mathematical DFA definition and a practical use case of them, namely testing if an input string is accepted.


\section{Related Work}

We discuss related work about visualizations of FSMs and related topics in ToC courses. 
JFLAP \cite{JFLAP} is a Java GUI that helps students visualize FSMs and some conversions provided students know some parts of their definitions and behavior.
FSMIPR uses a formal definition written in Python to generate the visualization, which requires more formal knowledge than JFLAP uses. 
In addition, FSMIPR is less interactive than JFLAP, with the intention of explaining concepts to students as opposed to practicing the less formal process of drawing out individual states. 
FSMIPR's visualizations are \emph{automatic} and thus are programmable, whereas JFLAP's are not. 

As far as we are aware, there are three relevant articles about FSM visualization.
The first is by Smith \cite{Smith2021}, with his motivation based on the limited resources available to undergraduate students when learning ToC concepts.
This project is limited to FSMs based on a regular expression.
In contrast, FSMIPR does not require knowing the regular expression in advance, and produces a video instead of a single static figure. 
Fransson's undergraduate thesis \cite{Fransson2013} evaluates a series of simulators in ToC and determines the one most suited for use based on functionality, documentation, and more.
They determined that JFLAP best met their needs due to feature availability and performance.
Singh's thesis \cite{Singh2007} describes simulations of FSMs and Pumping Lemma visualization, several of which are part of a ``hypertextbook'' with online animations and videos.
In contrast, FSMIPR is programmable and thus can be used both by students and teachers on new FSMs not statically given in a textbook.

\section{FSMIPR Structure and Results}

\begin{figure*}
    \centering
    \includegraphics[width=0.7\linewidth]{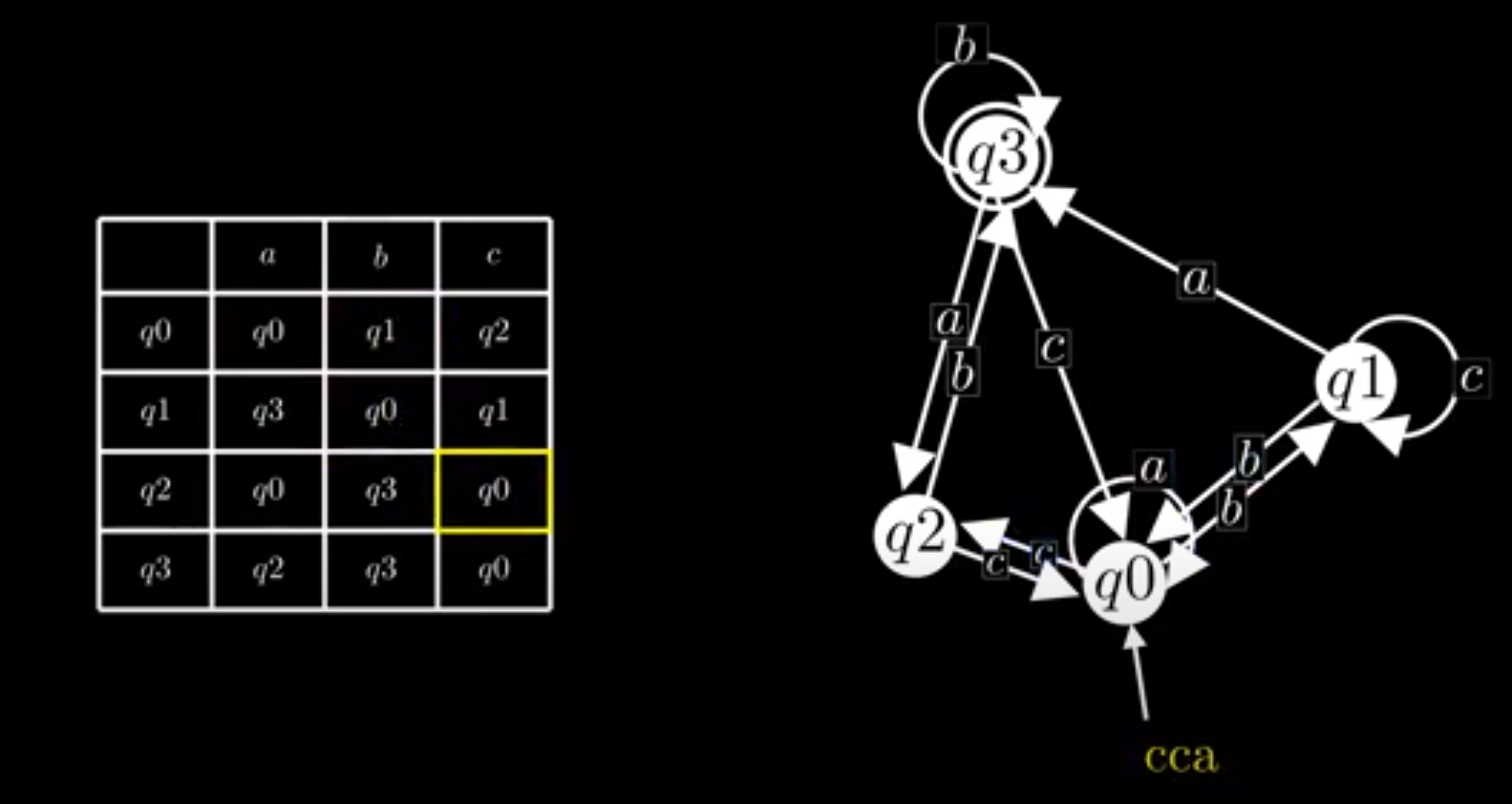}
    \caption{An example FSM automatically generated using FSMIPR.}
    \label{fig:example_fsm}
\end{figure*}

We developed FSMIPR in Python using the Manim \cite{Manim} and Automata-lib \cite{Automata} Python packages. 
Manim is an open-source math visualization and animation engine that outputs video, and Automata-lib contains implementations for FSMs found within ToC courses and provides methods that ``step'' through the FSM when running on an input.
To use the FSMIPR tool, all that is needed is a setup script, which must create an FSM object using Automata-lib and according to its formal definition, and additionally have a valid input string for the FSM.
After a method call to FSMIPR, running this script will automatically generate a video of this FSM's simulation on that input.\footnote{An example simulation video is: \url{https://www.youtube.com/watch?v=C07KhvqrbrI}.}
We use GraphViz \cite{graphviz} to automatically lay out the states.
For self-loop transitions, we attempt as best as possible to avoid crossing with other transitions and states.

The advantage of this tool over static images is that all relevant animations within a single step occur \emph{simultaneously}. 
Observe the FSM in Figure~\ref{fig:example_fsm}; the current next character to read in the input string is {\tt c}.
The left table is the FSM's transition table, and the FSM is currently in state $q2$.
After reading a {\tt c}, the FSM will then be in state $q0$.
Three events occur simultaneously: the {\tt c} is removed from the input string at the bottom, an animation of the $q_2 \xrightarrow{\tt c} q_0$ transition will be highlighted, and the transition table's entry for the $q_2$ row and column for the character {\tt c} will be highlighted.
Common issues for students in ToC courses when simulating FSMs is managing multiple pieces of data, and traditionally they will do so by iteratively updating one datum at a time.
We believe that FSMIPR can partially alleviate some of these issues.

\section{Conclusions and Future Work}

We created FSMIPR, a software package that provides automatic animations for running finite state machines when given an input string.
This tool is immediately useful for both ToC educators and students. 

We describe currently in-progress additions to FSMIPR.
In the Spring 2025 offering of our ToC course, we will launch a qualitative study to determine student opinion on FSMIPR's helpfulness on their learning.
Next, we will add automatic visualizations for conversions, algorithms, other formally defined objects in ToC courses; examples include NFA to DFA, Regex to NFA, and DFA minimization.
We will also add visualizations for common proofs within ToC courses, such as structural induction, Pumping Lemma, and undecidability proofs. 
Finally, we are investigating ``chaining'' multiple visualizations together when useful, such as generating a regular expression without a specific substring; the simplest method is first making a DFA and converting that to a regular expression.
\begin{acks}
The opinions in this work are solely of the author, and do not necessarily reflect those of the U.S. Army, U.S. Army Research Labs, the U.S. Military Academy, or the Department of Defense.  
\end{acks}

\bibliographystyle{ACM-Reference-Format}
\bibliography{sample-base}

\end{document}